\documentclass{vgtc}                          

\ifpdf
  \pdfoutput=1\relax                   
  \usepackage{graphicx}                
  \DeclareGraphicsExtensions{.pdf,.png,.jpg,.jpeg} 
\else
  \usepackage{graphicx}                
  \DeclareGraphicsExtensions{.eps}     
\fi%

\graphicspath{{figures/}{pictures/}{images/}{./}} 

\usepackage{microtype}                 
\PassOptionsToPackage{warn}{textcomp}  
\usepackage{textcomp}                  
\usepackage{mathptmx}                  
\usepackage{times}                     
\usepackage{cite}                      
\usepackage{tabu}                      
\usepackage{booktabs}                  

\usepackage{subfig}
\usepackage{amsmath}

\onlineid{0}

\vgtccategory{Research}

\vgtcinsertpkg

\title{STRIELAD - A Scalable Toolkit for Real-time Interactive Exploration of Large Atmospheric Datasets}

\author{Simon Schneegans\textsuperscript{1} \thanks{e-mail: simon.schneegans@dlr.de} %
\and Lori Neary\textsuperscript{2} \thanks{e-mail: lori.neary@aeronomie.be} %
\and Markus Flatken\textsuperscript{1} \thanks{e-mail: markus.flatken@dlr.de} %
\and Andreas Gerndt\textsuperscript{1} \thanks{e-mail: andreas.gerndt@dlr.de}}
\affiliation{\scriptsize \textsuperscript{1}German Aerospace Center (DLR), Dept. Simulation and Software Technology, \textsuperscript{2}Institut royal d'Aéronomie Spatiale de Belgique}

\teaser{
  \includegraphics[width=\textwidth]{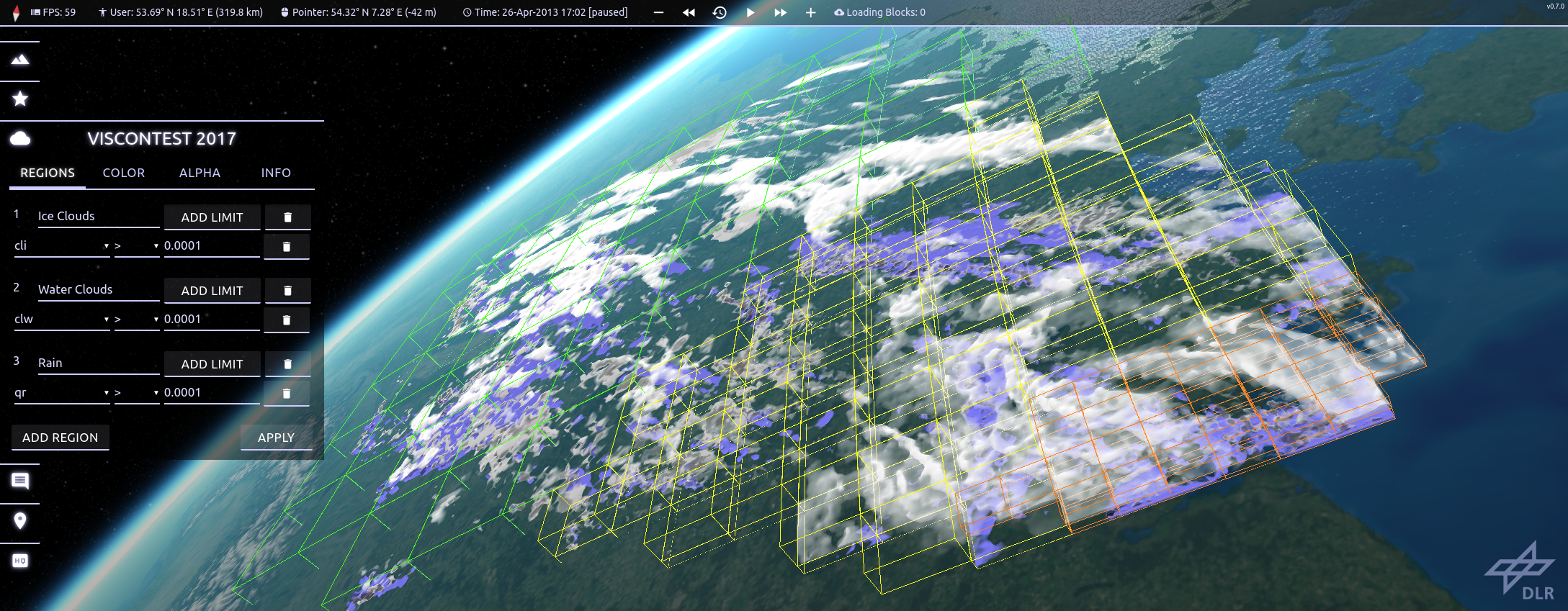}
  \caption{Real-time volumetric visualization of ice, water, and rain bands using our sub-volume selection mechanism in the graphical user interface. In addition, large terrain data is combined with the weather simulation to analyze their correlations.}
	\label{fig:teaser}
}

\abstract{
Technological advances in high performance computing and maturing physical models allow scientists to simulate weather and climate evolutions with an increasing accuracy. While this improved accuracy allows us to explore complex dynamical interactions within such physical systems, inconceivable a few years ago, it also results in grand challenges regarding the data visualization and analytics process.

We present STRIELAD, a scalable weather analytics toolkit, which allows for interactive exploration and real-time visualization of such large scale datasets. It combines 
parallel and distributed feature extraction using high-performance computing resources with smart level-of-detail rendering methods to assure interactivity during the complete analysis process.
} 

\CCScatlist{ 
  \CCScat{C.2.4}{Distributed Systems}{Client/Server}{Distributed applications};
  \CCScat{I.3.2}{Graphics Systems}{Distributed/network graphics}{Remote systems};
	\CCScat{I.6.6}{Simulation Output Analysis};
	\CCScat{J.2}{Physical sciences and engineering}{Earth and atmospheric sciences}
	 
}

\begin{document}

\firstsection{Motivation}
\maketitle


The substantial progress in high-performance computing (HPC) together with an improved understanding of natural processes enable scientists to model and simulate physical phenomena with an increasing accuracy and complexity. A major drawback of these technological advances is the growing gap between the possibilities to simulate and to analyze the produced data. It is not only a challenge of data complexity; the sheer amount of data, generated e.g. by numerical simulations, require tremendous effort to enable efficient data analysis and knowledge discovery. Especially interactive and explorative approaches are promising but also challenging. Yet they are important for urgent decision making in various fields such as traffic- or catastrophe management. Only when these gaps are closed, the full potential for knowledge discovery can be exploited.

STRIELAD, a scalable toolkit for real-time interactive exploration of large atmospheric datasets, will enable end-users to explore such large volumes of simulation data in real-time. It combines smart data acceleration structures, parallel and distributed processing, innovative data selection and interaction mechanisms. This is demonstrated with the 2.9~TB weather simulation dataset provided for the 2017 IEEE visualization contest. Additionally, this data is combined with various other data sources such as flight trajectories and geographical information obtained from satellite imagery using DLR's terrain rendering software \cite{westerteiger2012spherical}. The goal of this paper is not to answer all given questions of the visualization contest. It is rather a description of STRIELAD, which in turn can be used to find those answers by real-time exploration.

\section{Scalable Visualization Toolkit}

A lot of applications for scientific visualization of simulation data are available on market. Some of them are scalable to enable the visualization of large-scale data~\cite{AHRENS2005717}~\cite{HPV:VisIt}. Furthermore, these applications often provide an extensive feature set for various application domains. However, to our knowledge none of them enables spatio-temporal exploration of huge  simulation data with interactive frame rates and feature extraction in range of milliseconds.

With STRIELAD, end-users are able to interactively select sub-volumes of a dataset and apply boolean operations on them. For each sub-volume, several scalar thresholds or domain limits can be defined.  To speed-up feature extraction, the processing is importance driven, distributed and highly parallel. Combined with level-of-detail rendering and immediate streaming of result data, this creates a very responsive and interactive system. A shader editor in the graphical user interface (GUI) allows to control the appearance of the defined sub-volumes. The shading possibilities range from informative to realistic.

\begin{figure}[tb]
  \centering 
  \includegraphics[width=\columnwidth]{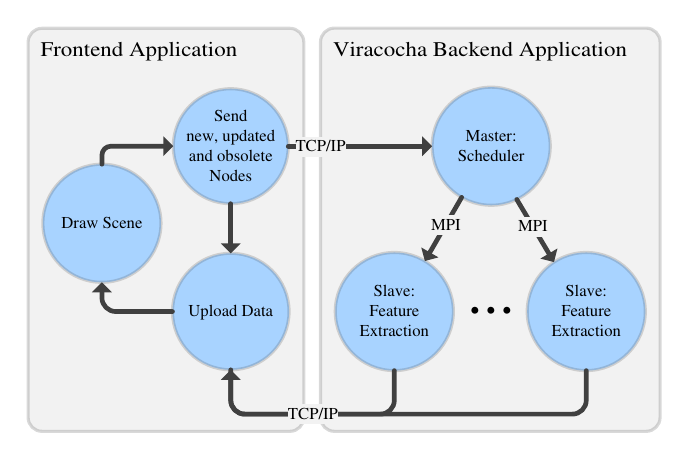}
  \caption{Distributed visualization setup: The frontend determines nodes for parallel processing. The backend executes the feature extraction on a HPC cluster system and streams results back.}
  \label{fig:viracocha}
\end{figure}

Figure~\ref{fig:viracocha} depicts the high-level architecture of the framework. A frontend application performs real-time rendering and user interaction. The GUI enables end-users to define multiple volumes of interests (depicted in \autoref{fig:teaser}). These sub-volumes are specified by several lower and upper scalar limits. Based on this information, the user's position and the simulation time, requests are compiled and sent to a backend application. There, feature extraction is performed on an octree which has been generated in a pre-processing step. Furthermore, the user has the ability to modify shader code at runtime via the GUI. This allows for powerful coloring and shading of the received geometry.

\subsection{Preprocessing}

Before data can be explored with STRIELAD, a preprocessing step generates the required acceleration structure. In this step, the rectlinear input dataset is partitioned into many equally sized blocks (leaf nodes), and an octree is generated by sub-sampling (see \autoref{fig:octree}). The additional memory requirement for the inner nodes of the octree is approximately 15~\% of the size of the original data. This is about 400~GB for the visualization contest dataset.

Since several terabytes of data are read and written during the preprocessing, the actual execution time strongly depends on disc speed and may take several hours.

While the size of the individual octree nodes has no influence on the memory requirements of the acceleration structure, it significantly influences the visualization performance. If the size is small, many nodes are generated. As a consequence, feature extraction in the backend is highly parallel and visualization updates are received with low latency. However, since many nodes have to be rendered by the frontend, a lot of draw calls have to be issued and therefore the communication with the graphics driver may become a bottleneck.

The ideal node size depends on disc speed, backend processing power, network speed, CPU and GPU performance of the frontend. We achieved very good results with a node size of $ 20 \times 20 \times 20 $ cells. This results in seven octree levels and a total node count of 51.393 per time step.

\subsection{Parallel feature extraction}

The parallel and distributed feature extraction application is based on Viracocha~\cite{Gerndt:2004:VR}. As depicted in \autoref{fig:viracocha}, our application implements a master-slave parallelization approach. The master process receives user queries from a frontend application, dissipates the requests into small independent work items, and distributes these work items to $n$ slave instances for parallel processing. Furthermore, an integrated caching approach increases the I/O bandwidth for repeated data processing. We enhanced the scheduling approach of Viracocha to update the processing order of a running computation. This allows to fully change the processing order, e.g. discard previously visible data, insert new items for processing, or rearrange the data processing based on priority values.

The slave processes execute a boolean intersection algorithm for sub-volume extraction. The algorithm uses VTK~\cite{Kitware:2003} to load and process the individual octree nodes. It gets a description of pipelined intersection $\cap$ operations defined by the limits per sub-volume. First, the dataset is clipped according to each limit, then the result is triangulated and normals are generated. Finally, it is transferred back to the frontend application for rendering. Since multiple sub-volumes can be defined on the frontend, union $\cup$ operations are also possible.

\subsection{Level-of-detail rendering}

\begin{figure}[tb]
  \centering
  \includegraphics[width=\columnwidth]{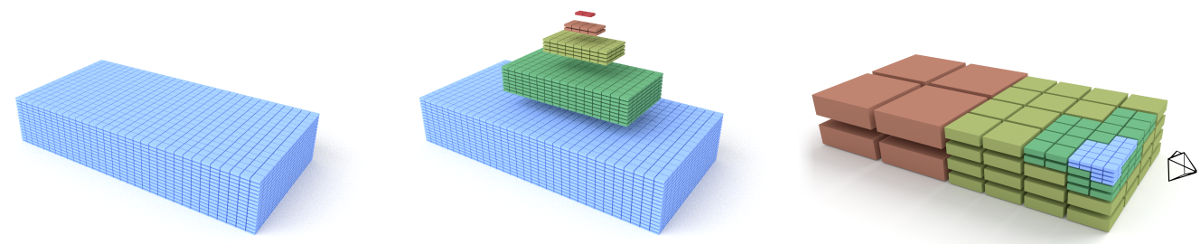}
  \caption{The original dataset is partitioned into many small data blocks (left). Then, by combining eight blocks into one, an octree is generated (center). When rendering, a set of nodes (called the `cut') is chosen so that blocks close to the virtual camera have a high resolution (right).}
  \label{fig:octree}
\end{figure}

The frontend application has two main tasks. On the one hand it has to maintain the `cut'; this is a set of nodes which are currently rendered on the screen (see \autoref{fig:octree}). For all changes of the cut, a message is sent to the backend. On the other hand, it has to perform the actual rendering of nodes for which data was received. \\

At first, only the root node is part of the cut. If it is visible and its bounding box covers a maximum solid angle, a split operation is performed: All visible children nodes are added to the cut and the root node is removed. This process continues recursively until all visible nodes have been refined according to the threshold angle.

Every frame this traversal is repeated, either adding nodes to the cut (because they became visible or due to a split operation) or removing nodes which either became invisible or which have been merged in a collapse operation. Each node added to the cut gets a custom priority, determined by the distance between the virtual camera and the center of the node's bounding box. Therefore, nodes close to the camera have a higher priority.

Then, three lists of nodes are assembled: One contains all newly visible nodes, one all nodes which are not visible anymore and one all nodes with changed priorities. These lists are sent to the backend which then starts the feature extraction for all new nodes, updates the priorities of scheduled requests and removes all obsolete requests.

When a parameter of the visualization is changed (e.g. the list of visualized sub-volumes or the current time step) all running requests are considered obsolete and for each node in the cut a completely new request is sent to the backend. The backend will then abort all pending requests. \\

Rendering of the visible nodes follows a simple paradigm: A node which becomes visible should always show up-to-date data (the currently configured sub-volumes and the correct time step). However, a node which has been visible before is allowed to show stale data until it either receives new data from the backend or it becomes invisible. This results in a desired characteristic: When the configured sub-volume or the time step are changed, the scene progressively updates, starting with the nodes close to the camera. When frequent visualization changes occur, previous tasks are aborted. Hence, if the backend is not fast enough, there will be frequent updates close to the camera and less updates to distant nodes.

Optionally, the wind field is used to move the sub-volumes between time steps. This helps to make the transitions between time steps much smoother and conveys the motion of clouds.

In order to allow for the desired pseudo-volumetric appearance, the extracted geometry is rendered to an A-Buffer \cite{Carpenter1984}. Usually, this data structure stores linked lists of semi-transparent fragments at each pixel which are filled during rasterization. In a final resolve step the lists are depth-sorted and alpha-composited. In STRIELAD however, no alpha value is written into the A-Buffer. Instead an unique sub-volume ID is stored for each fragment. This allows for resolving corresponding entry and exit points of arbitrarily intersecting sub-volumes along the view ray.

The color written into the A-Buffer is calculated by a user-defined function. This gets as input the sub-volume ID, all available interpolated scalar values at the corresponding point in the volume and some additional information such as the surface normal and the direction of the sun. A second function is invoked in the A-Buffer resolve step. It returns a transparency value based on the sub-volume ID and the distance between entry and exit points into the volume.

\section{Results}

\begin{figure}%
  \centering
  \subfloat[8 FPS]{{\includegraphics[width=4.1cm]{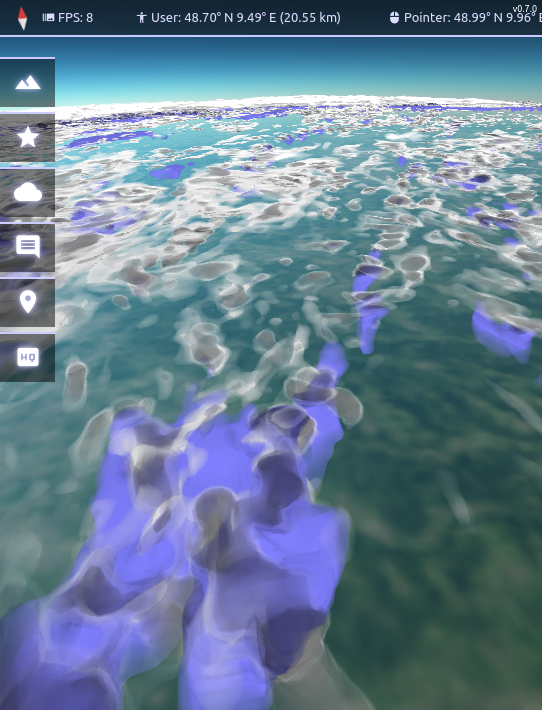} }}%
  \hfill
  \subfloat[67 FPS]{{\includegraphics[width=4.1cm]{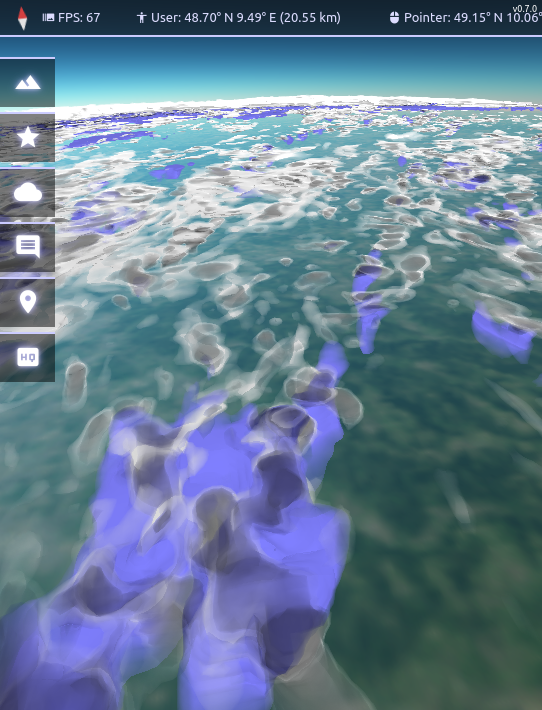} }}%
  \caption{The full dataset cannot be rendered at interactive frame rates (a). However, this is possible if a view dependent resolution is used in the distance. The error in the distance is barely noticeable.}%
  \label{fig:fps}%
\end{figure}

The developed toolkit can be used for real-time analysis of arbitrarily large structured atmospheric datasets in time and space.

With the level-of-detail rendering approach, all time steps of the entire dataset can be explored at interactive frame rates (see \autoref{fig:fps}). When stepping through time, updating the current perspective takes only a couple of seconds. Since nodes in the foreground will receive data in less than a second this results almost in a fluent animation. This is especially true if the wind field is used to move the clouds between time steps. Furthermore, as the scene is updated progressively, the user can move the camera during the update with the full frame rate.

The user can specify multiple sub-volumes bounded by arbitrary scalar thresholds. This can be used to identify volumes of interest such as global extrema, trends and correlations between different scalars. Complex boolean operations can be performed on the dataset. The user can specify color and opacity for all sub-volumes with a powerful shader editor which allows for many different scenarios. The color can be based on all available scalars, their respective minimum and maximum values, the surface normal, the sun direction, the position in 3D space and other parameters.

Additionally, various digital elevation models (e.g. SRTM30) can be used together with other GIS sources such as OpenStreetMap or the Bluemarble dataset by NASA. The position of Earth and Sun is calculated with the SPICE library \cite{spice} and represents the conditions at the simulation time. Finally, the flight trajectories provided for the visualization contest are shown as lines with a user defined length.

\section{Discussion and future work}

With STRIELAD, we presented a scalable and interactive visualization environment for the analysis of large time-dependent weather simulations in combination with geospatial data. A level-of-detail rendering and processing approach combined with distributed and parallel feature extraction enables the real-time exploration of data. However, by now the feature extraction is limited to the selection of sub-volumes using boolean operators with powerful styling capabilities due to the shader editor. The integration of further algorithms such as slicing or contouring will be straight forward. More challenging will be algorithms using global information such as topological features or tracing methods. For these algorithms several aspects of the parallelization and data distribution will need substantial changes. Furthermore, our approach produces some artifacts when the maximum number of fragments in the A-Buffer is reached. For this purpose different fragment inserting strategies are necessary.

In the near future, we plan to add interpolation methods to the pre-processing. This should lead to much smoother transitions between the resolution levels of the LOD data structure. Additionally, we will extent the application to run on our tiled virtual reality system. To further speedup the feature extraction, we will investigate in data prefetching strategies.


\bibliographystyle{abbrv-doi}

\bibliography{template}

\pagebreak

\begin{figure*}[tb]
  \centering
  \includegraphics[width=12cm]{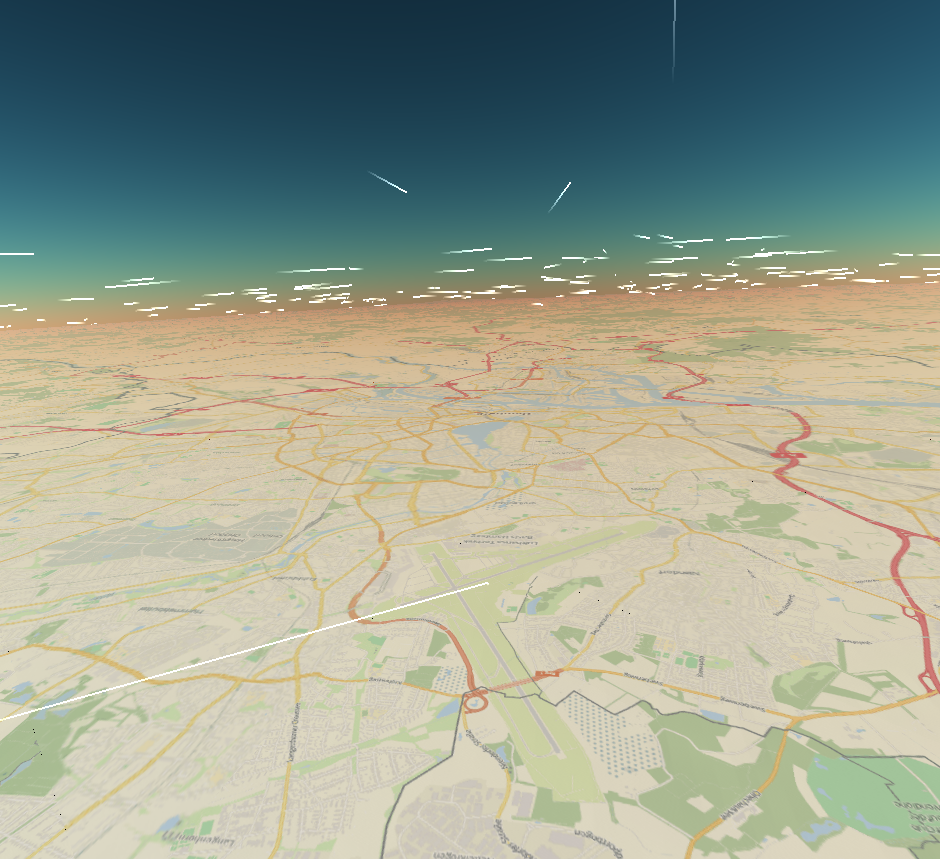}
  \caption{An airplane landing at Hamburg Airport. All 27000 airplanes are loaded an populate the scene with their trajectories. The length of the trajectories can be adjusted and defaults to the distance the respective airplane traveled in two minutes}
\end{figure*}

\begin{figure*}%
  \centering
  \subfloat[Visualization of color coded absolute wind speed in the south east of the domain. Interactions between the wind magnitude and the SRTM30 terrain are clearly visible.]{{\includegraphics[height=7cm]{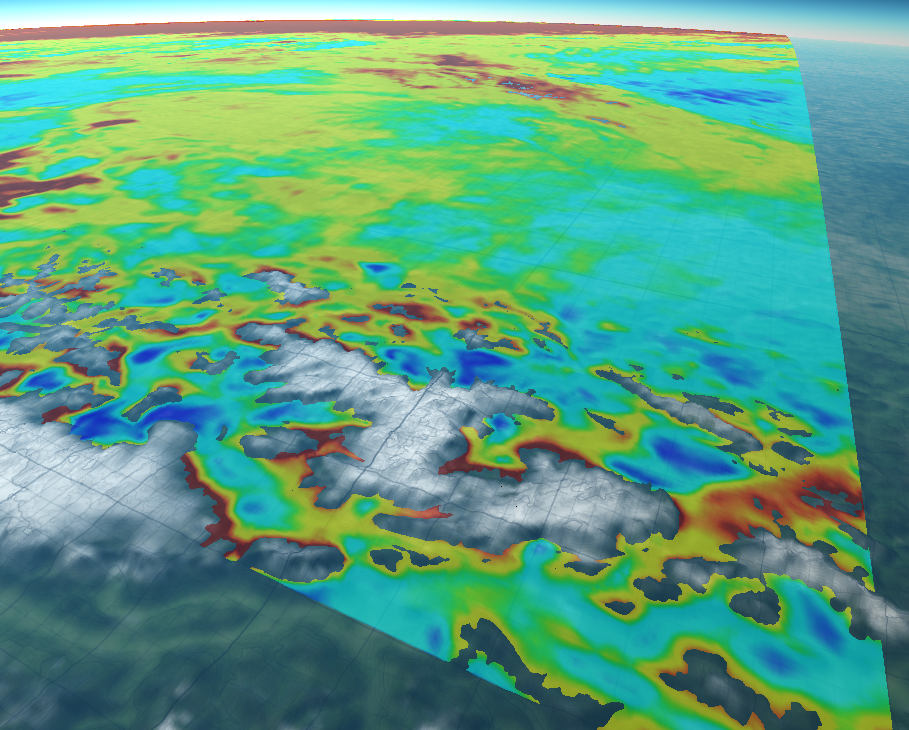} }}%
  \hfill
  \subfloat[Sub-volumes of high rain mixing ration color coded by the wind magnitude at their surfaces. Red areas indicate a lot of rain with high wind speeds.]{{\includegraphics[height=7cm]{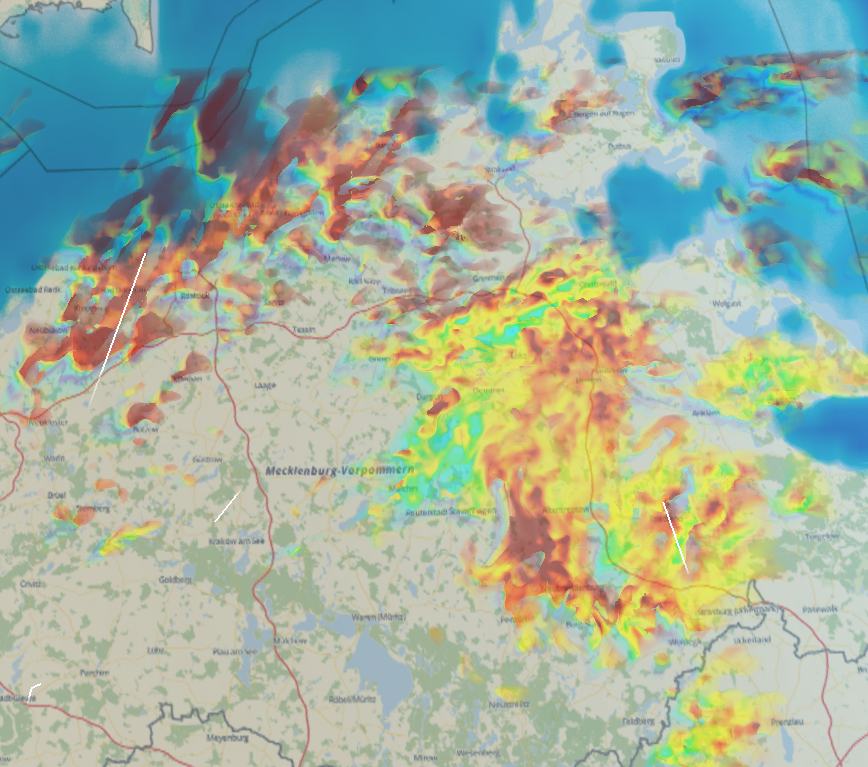} }}%
  \caption{The sub-volumes can be colored with complex shader code. In these examples scalars are mapped on a color scale.}%
\end{figure*}

\begin{figure*}%
  \centering
  \subfloat[]{{\includegraphics[width=5.5cm]{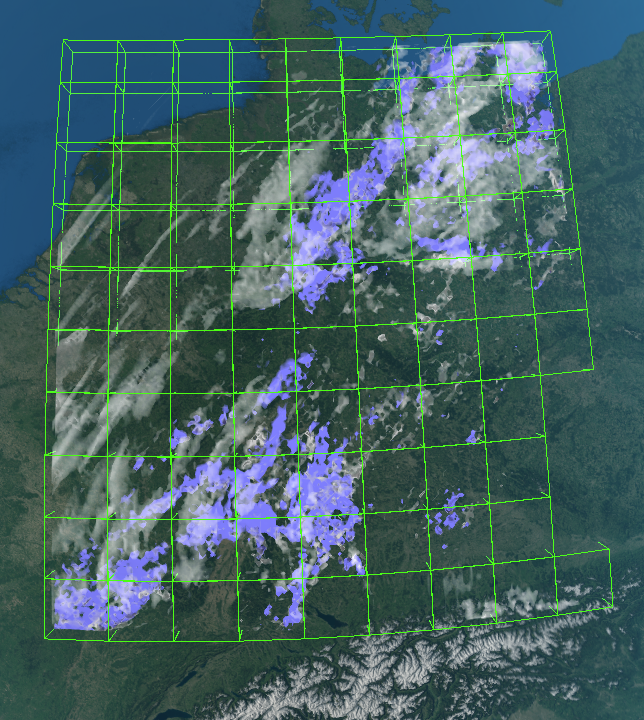} }}%
  \hfill
  \subfloat[]{{\includegraphics[width=5.5cm]{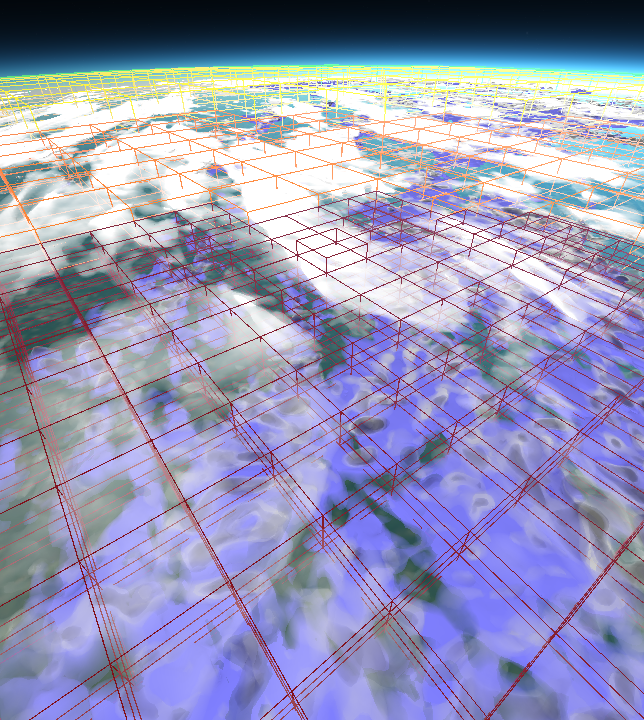} }}%
  \hfill
  \subfloat[]{{\includegraphics[width=5.5cm]{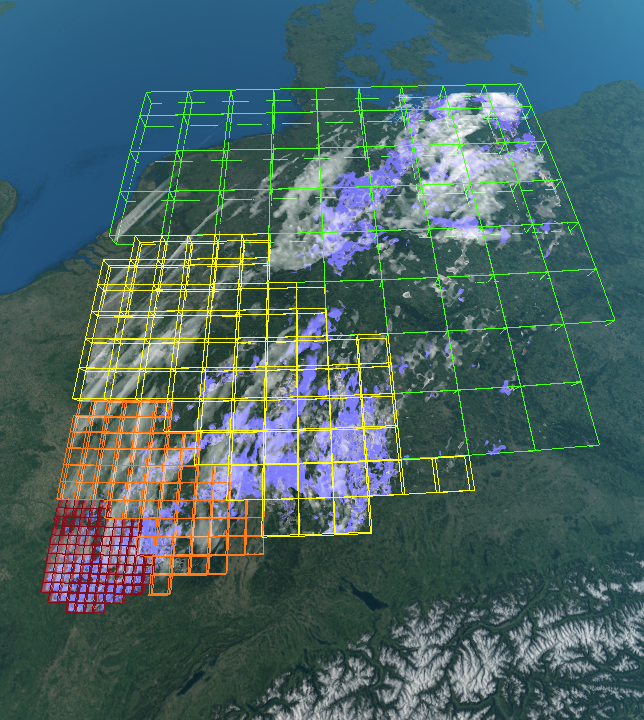} }}%
  \caption{When the dataset is rendered for perspective (a), medium resolution nodes of the octree are chosen (illustrated with the green bounding boxes). If the camera is moved closer to the dataset, nodes with better resolution are loaded close to the camera (b). Image (c) shows the same nodes as (b) but rendered from an outside perspective.}%
\end{figure*}

\begin{figure*}[tb]
  \centering
  \includegraphics[width=16cm]{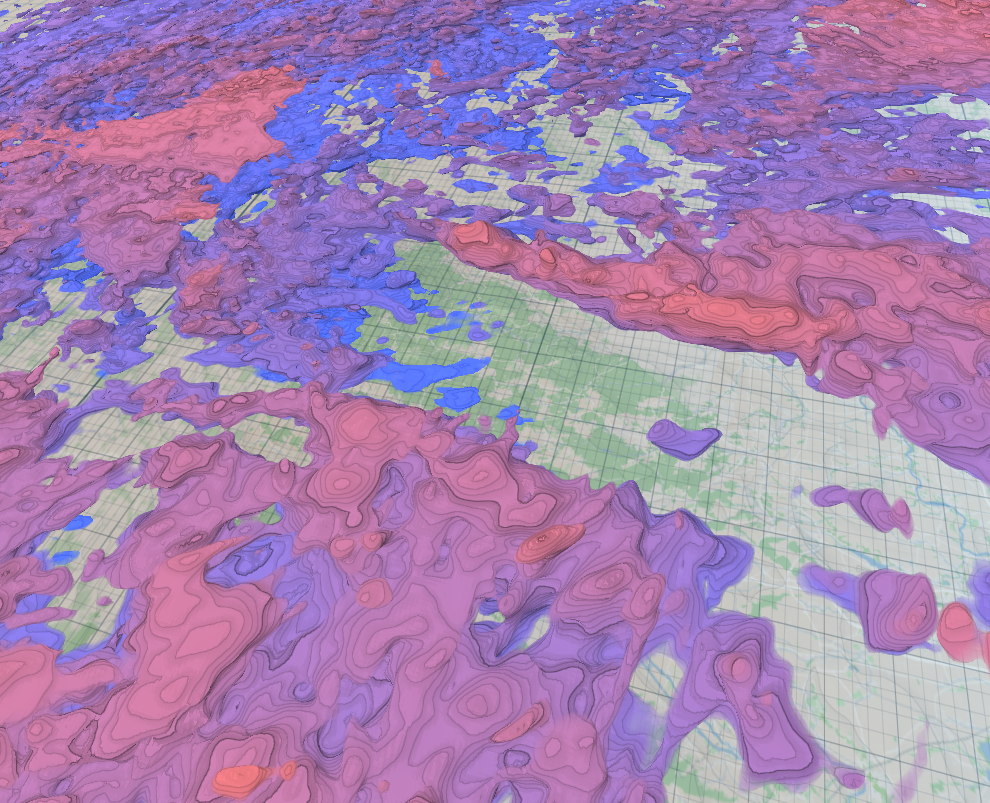}
  \caption{The shader editor allows expert users to achieve complex results. In this example, Sub-volumes of high cloud water content around the Harz are enriched with iso-altitude lines. Thin lines are spaced 100~m, thick lines 1~km in height. This helps to visualize the topology of the features.}
\end{figure*}

\begin{figure*}[tb]
  \centering
  \includegraphics[width=\textwidth]{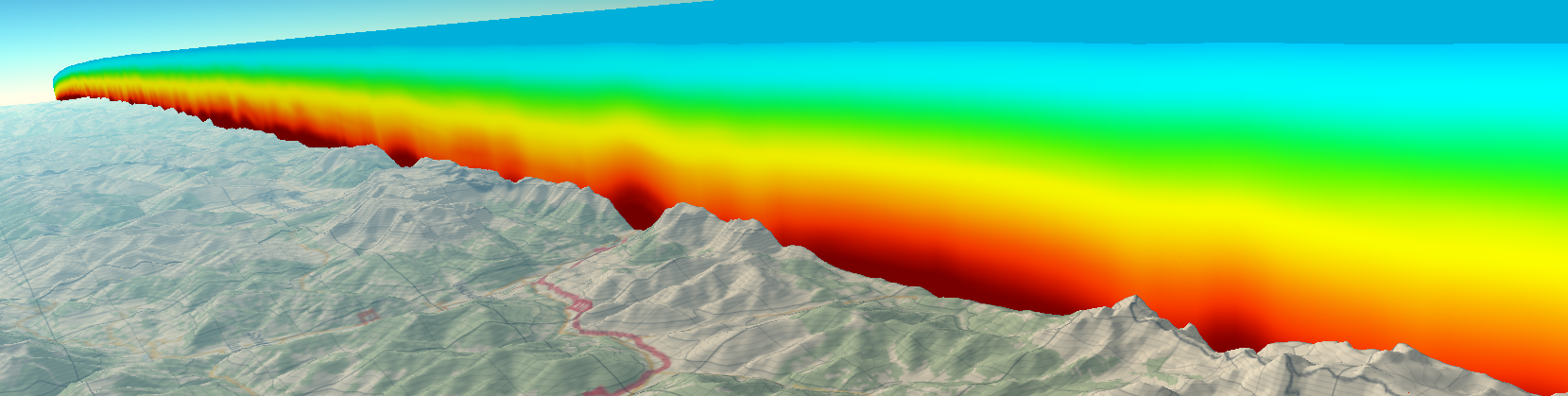}
  \caption{The southern border of the dataset with color-coded temperature shown with SRTM30 digital elevation model and OpenStreetMap data. The correlation between terrain and temperature distribution is apparent.}
\end{figure*}

\begin{figure*}[tb]
  \centering
  \includegraphics[width=\textwidth]{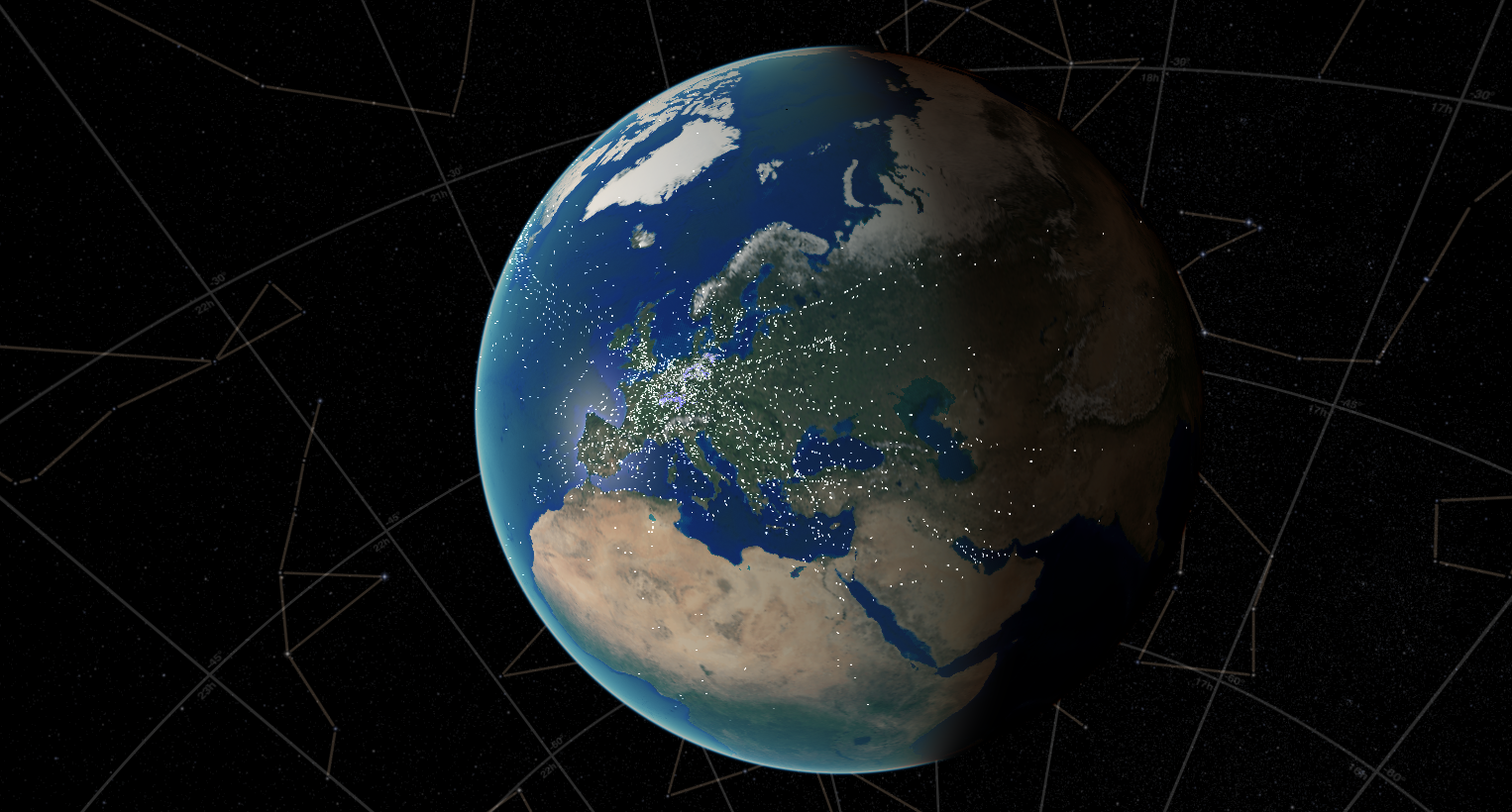}
  \caption{The visualization framework is integrated in an OpenGL application for rendering entire planetary datasets. In the screen shot above the relative position of Earth and Sun are as they were on 26/04/2013 at 5:03 PM. The tiny white dots are the airplane trajectories which were provided for the 2017 IEEE visualization contest.}
\end{figure*}

\begin{figure*}[tb]
  \centering
  \includegraphics[width=\textwidth]{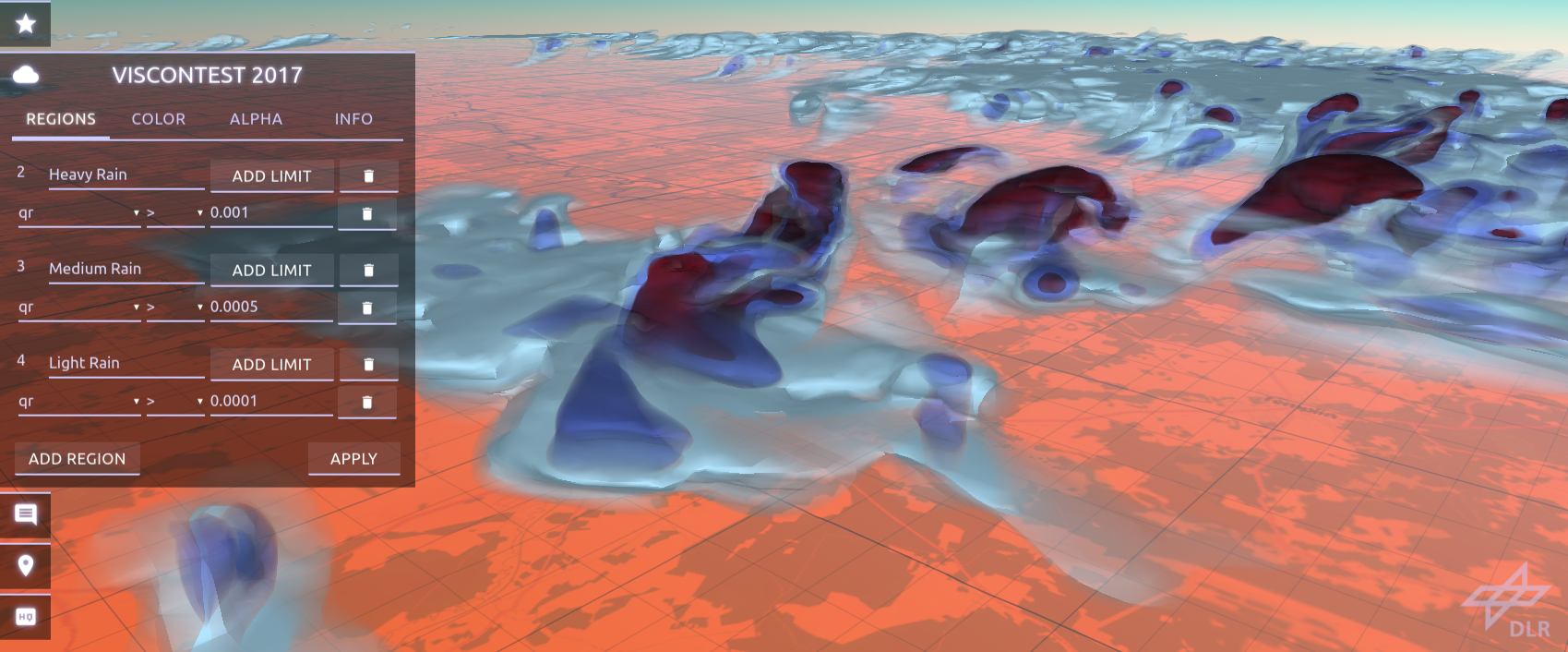}
  \caption{This is an example of intersecting volumes. Sub-volumes for three rain mixing ratios are inside each other. The pseudo-volumetric effect helps to understand the depth of the extracted regions.}
\end{figure*}

\begin{figure*}[tb]
  \centering
  \includegraphics[width=\textwidth]{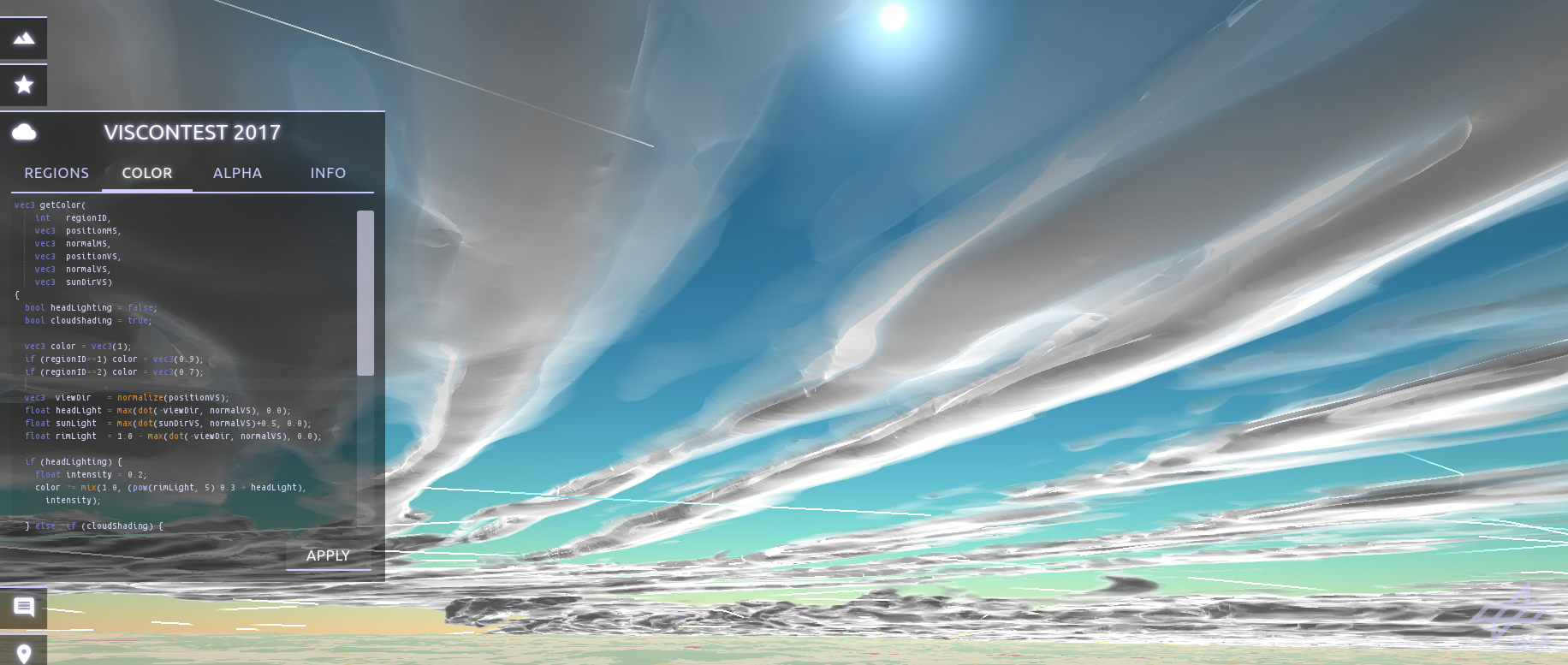}
  \caption{High altitude clouds in the souther part of the simulation domain. On the left hand side the shader editor in the user interface can be seen. There are also some airplane trajectories visible in this image.}
\end{figure*}

\begin{figure*}[tb]
  \centering
  \includegraphics[width=\textwidth]{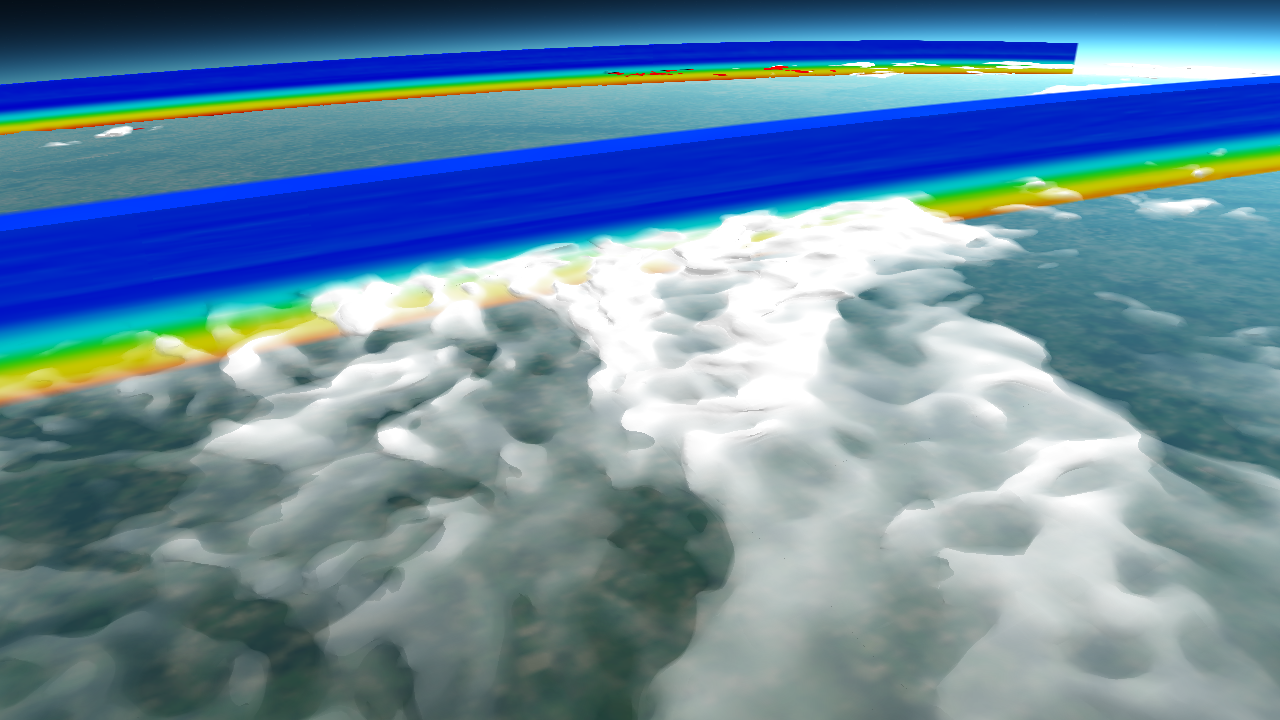}
  \caption{This visualization shows two slices of the dataset color-coded according to temperature. In addition, high altitude ice clouds are shown in white. Non of the ice clouds extends into the tropopause (the darkest blue indicates the coldest layers of air) which would indicated a severe thunderstorm.}
\end{figure*}

\begin{figure*}[tb]
  \centering
  \includegraphics[width=\textwidth]{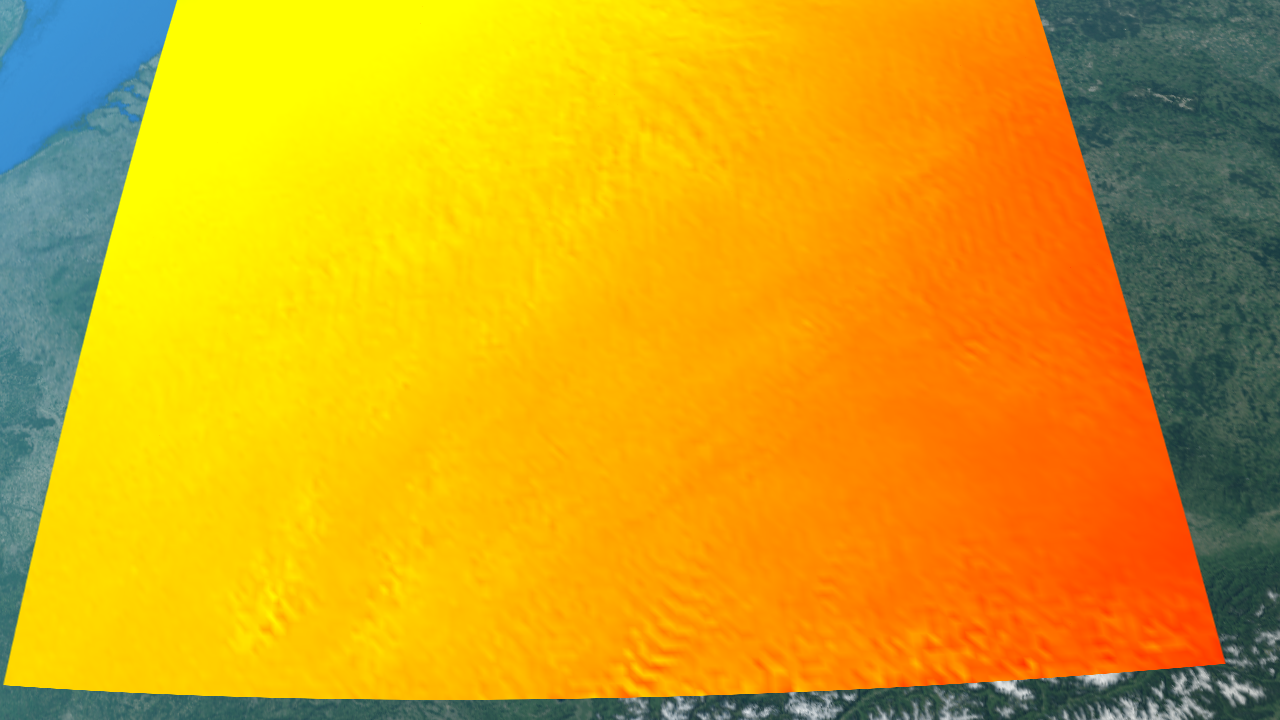}
  \caption{If a slice of the volume is color-coded according to pressure, gravity waves become visible.}
\end{figure*}

\end{document}